\documentclass[manuscript]{acmart}
\AtBeginDocument{%
  \providecommand\BibTeX{{%
    \normalfont B\kern-0.5em{\scshape i\kern-0.25em b}\kern-0.8em\TeX}}}

\copyrightyear{2021}
\acmYear{2021}
\setcopyright{rightsretained}
\acmConference[AutomotiveUI '21 Adjunct]{13th International Conference on Automotive User Interfaces and Interactive Vehicular Applications}{September 9--14, 2021}{Leeds, United Kingdom}
\acmBooktitle{13th International Conference on Automotive User Interfaces and Interactive Vehicular Applications (AutomotiveUI '21 Adjunct), September 9--14, 2021, Leeds, United Kingdom}\acmDOI{10.1145/3473682.3479717}
\acmISBN{978-1-4503-8641-8/21/09}

\begin{document}

\title[CUI @ Auto-UI]{CUI @ Auto-UI: Exploring the Fortunate and Unfortunate Futures of Conversational Automotive User Interfaces}

\author{Justin Edwards}
\affiliation{%
  \institution{University College Dublin}
  \city{Dublin}
  \country{Ireland}}
\email{justin.edwards@ucdconnect.ie}

\author{Philipp Wintersberger}
\affiliation{%
  \institution{Technische Universität Wien}
  \city{Vienna}
  \country{Austria}}
\email{philipp.wintersberger@tuwien.ac.at>}

\author{Leigh Clark}
\affiliation{%
  \institution{Swansea University}
  \city{Swansea}
  \country{UK}}
\email{l.m.h.clark@swansea.ac.uk}

\author{Daniel Rough}
\affiliation{%
  \institution{University of Dundee}
  \city{Dundee}
  \country{UK}}
\email{drough001@dundee.ac.uk}

\author{Philip R. Doyle}
\affiliation{%
  \institution{University College Dublin}
  \city{Dublin}
  \country{Ireland}}
\email{philip.doyle1@ucdconnect.ie}

\author{Victoria Banks}
\affiliation{%
  \institution{University of Southampton}
  \city{Southampton}
  \country{UK}}
\email{v.banks@soton.ac.uk}

\author{Adam Wyner}
\affiliation{%
  \institution{Swansea University}
  \city{Swansea}
  \country{UK}}
\email{a.z.wyner@swansea.ac.uk}

\author{Christian P. Janssen}
\affiliation{%
  \institution{Utrecht University}
  \city{Utrecht}
  \country{The Netherlands}}
\email{C.P.Janssen@uu.nl}

\author{Benjamin R. Cowan}
\affiliation{%
  \institution{University College Dublin}
  \city{Dublin}
  \country{Ireland}}
\email{benjamin.cowan@ucd.ie}

\renewcommand{\shortauthors}{Edwards, et al.}

\begin{abstract}
  This work aims to connect the Automotive User Interfaces (Auto-UI) and Conversational User Interfaces (CUI) communities through discussion of their shared view of the future of automotive conversational user interfaces. The workshop aims to encourage creative consideration of optimistic and pessimistic futures, encouraging attendees to explore the opportunities and barriers that lie ahead through a game. Considerations of the future will be mapped out in greater detail through the drafting of research agendas, by which attendees will get to know each other's expertise and networks of resources. The two day workshop, consisting of two 90-minute sessions, will facilitate greater communication and collaboration between these communities, connecting researchers to work together to influence the futures they imagine in the workshop.
\end{abstract}




\maketitle

\section{Introduction}
Over the last several years, there has been significant interest in the use of conversational interfaces in cars, with several publications on the topic appearing in the Automotive User Interfaces (Auto-UI) and Conversational User Interfaces (CUI) conferences \cite{funk_usable_2020,large_its_2019,scatturin_left_2020,stier_towards_2020-1,schartmuller2019text,frison2019athena}. Despite this overlap in interest, there is yet little collaboration and communication between the Auto-UI and CUI communities. This workshop aims to bring the two communities together through an open-minded discussion of the future of automotive conversational user interfaces, discussing the opportunities as well as the challenges.  

Underpinning the work already underway at the intersection of 
CUI and Auto-UI, these communities share interests in multimodal interaction evaluation \cite{vo_investigating_2020, jaber_patterns_2019,wu_mental_2020}, multitasking and interruptions as interaction paradigms \cite{edwards_multitasking_2019,wong_voices_2019,funk_usable_2020,landesberger_what_2020,wintersberger2018let}, modeling mental workload \cite{wu_mental_2020, pakdamanian_toward_2020, funk_usable_2020}, and mixed-methods approaches to research ranging from physiological sensing \cite{pakdamanian_toward_2020,glatz_use_2018,jaber_patterns_2019} to in-the-wild observation \cite{fischer_progressivity_2019,braun_automotive_2018}. We aim to bring together the shared goals and compare the different approaches of these communities, establishing a community of practice that can share resources and expertise to better understand automotive conversational user interfaces.

\section{Workshop Activities and Goals}
The workshop primarily aims to encourage wide-ranging discussions about the future of automotive conversational user interfaces in order to promote future research partnerships between the Auto-UI and CUI communities, following in the tradition of previous "CUI @" workshops \cite{murad_lets_2021,porcheron_cuicscw_2020,candello_cuichi_2020,doyle_cuiiui_2021}. To encourage creative, divergent consideration of the future of this topic, this workshop aims to gamify conversations, drawing inspiration from the "What Could Go Wrong" card game from a previous Auto-UI workshop \cite{martelaro_what_2020}. We plan to use the broad discussions from an initial session to get to know the workshop attendees and their views of the future of conversation in the car. This will be used to tailor a focused second session aimed at mapping research agendas by which we can understand, seek, or avoid those imagined futures. Because we see concrete research collaboration as the ultimate goal of this workshop, the second session will feature focused methodological discussion to help to make attendees aware not only of each other's interests, but also their expertise. The workshop organizers will leverage the existing communication channels established through past workshops to better connect attendees to researchers in similar topics that primarily see their research home in other conferences.

\subsection{"Fortunately/Unfortunately" Game}
To maximize our time spent meaningfully discussing the future of conversational automotive user interfaces, the workshop will begin with attendees getting randomly assigned to small groups to play a game of "Fortunately/Unfortunately". By making the first activity of the workshop a discussion-based game, we hope to establish a creative and lighthearted mood early in the workshop, forgoing a need for icebreakers or introductions. This game format is inspired by a previous Auto-UI workshop \cite{martelaro_what_2020} which used a Games with a Purpose \cite{von_ahn_designing_2008} approach to facilitate discussion. The "What Could Go Wrong?" game featured in that workshop \cite{martelaro_what_2020} and the Game with a Purpose "Judgement Call" \cite{ballard_judgment_2019} both aimed to make it easier for players to discuss difficult topics regarding downsides and ethical dilemmas resulting from technology. We chose a game format for just the same reason - allowing, and indeed encouraging, attendees to consider the challenges and dark sides of a conversational automotive interface future, alongside the more positive opportunities.

Fortunately/Unfortunately is an improvisational, discussion-based game in which players take turns telling a story or describing a scenario line-by-line, alternating whether they begin their turn with the word "fortunately" or the word "unfortunately". This format pushes players to think creatively, dreaming up potential risks or downsides for even the most optimistic scenarios and considering solutions and silver linings when presented with negativity. A full round of the game is complete when each player has contributed one line to a scenario. Each group will have one or more workshop organizers in the group to write down each line of the scenario, and after a round of the game, all attendees will be encouraged to talk through each line, expanding on the upsides or downsides from each twist in the game. These discussions will be visually mapped on interactive Miro boards\footnote{www.miro.com} so organizers and attendees can consider the branching consequences of future decisions. 

\subsection{Mapping Research Agendas}
In the second session of the conference, organizers will select scenarios from the first session and create small groups of attendees to map a research agenda
based on a given scenario. Scenarios will be selected to represent a variety of areas of research, such as legal and safety challenges, the future of work, and technological challenges. Groups will be assigned to ensure each group has a diversity of expertise and that group members' research interests align with the scenario they are assigned to. Each group, along with one or more organizers, will be asked to draft a research agenda that addresses questions raised at each line from the scenario, exploring the methodological approaches they might consider employing to answer those research questions. Groups will be encouraged to pretend they have won a large grant to work on these questions over the next five years, with research expenses covered and the opportunity to work with colleagues who aren't present in their group. This frame will push participants to share their own expertise and to talk through the professional networks available to them, discussing the human and research capital within their network. Workshop organizers will again map these discussions visually on Miro boards. 

\section{Tentative Schedule and Engagement}
Before the workshop, we plan to engage attendees by asking them to submit a short research biography as well as 3 optimistic views and 3 pessimistic views regarding the future of automotive conversational user interfaces. These views will be collected by organizers to use as potential starting points for Fortunately/Unfortunately game rounds. The workshop will be split across two 90-minute sessions across two days.  The schedule of the first day will include:
\begin{itemize}
  \item Introductory remarks (10 minutes): Organizers will briefly introduce themselves and the aims of the workshop and give a brief demonstration of a round of Fortunately/Unfortunately.
  \item Fortunately/Unfortunately games (50 minutes): Organizers will create breakout rooms of 4-7 attendees each, facilitating games of Fortunately/Unfortunately. Organizers will create Miro board depictions of each round of the game, asking participants to expand on details of their envisioned scenarios, and collecting these thoughts on the Miro boards as well. Multiple rounds of Fortunately/Unfortunately with discussions should take place across this section of the workshop.  
  \item Group discussion of game scenarios (30 minutes): After playing games of  Fortunately/Unfortunately in small groups, we will come back together as a whole to discuss the scenarios each group laid out. Organizers will share the Miro boards they created to walk the group at large through the creative process that led to each vision of the future.
\end{itemize}

The second day's schedule will include: 
\begin{itemize}
  \item Introductory remarks (10 minutes): Organizers will reintroduce the aims of the workshop and explain the framing of the research agenda mapping session.
  \item Small group research agenda mapping (50 minutes) Organizers will again create breakout rooms, with group membership curated by organizers between sessions based on research bios submitted before the workshop, to plan a five-year research agenda based on a single scenario from the first session. Organizers will create new Miro boards with information from the scenario's original board copied over, so ideas can be mapped out and saved in a persistent document.
  \item Group discussion of research agendas (30 minutes): After planning research agendas in small groups, the workshop will come together as a whole again to discuss the research agendas they planned, the publication venues they would hope to target with the work they imagined, and shared visions of the future. A final Miro board will be created to document the group-wide discussion and participants will receive information for joining an existing CUI mailing list and Slack workspace with a channel created for automotive research. 
\end{itemize}

\section{Planned Outcomes}
Following the workshop, we aim to facilitate further collaboration between the CUI and Auto-UI communities in the following ways:
\begin{enumerate}
  \item Sharing the research agendas drafted on the second day of the workshop, publishing both the Miro boards and executive summaries of each research agenda to all workshop attendees. These summaries will facilitate collaboration by attaching workshop attendees to their areas of interest and methodological expertise in a single persistent record available on the CUI community website (https://www.conversationaluserinterfaces.org)
  \item Connecting workshop attendees to the broader CUI research community by encouraging them to join the established mailing list and Slack workspace, with a new channel for automotive CUI research.
  \item Identifying venues for future collaboration opportunities, including the CUI conference provocation papers or posters tracks, the Auto-UI Late Breaking Works or videos tracks, CHI Late Breaking Works and alt.CHI tracks as well as potential future workshops at venues like CSCW, IUI, CHI, and MobileHCI.
\end{enumerate}

\section{Organizers}
{\textbf{Justin Edwards}} is a PhD candidate at University College Dublin’s School of Information \& Communication Studies. His research focuses on the design of proactive speech-based agents, the cognitive implications of multitasking with speech, and understanding of spoken interruptions. He has served on the organizing committee of all three CUI Conferences and acted as a workshop organizer for multiple "CUI @" workshops (e.g. IUI and CHI)

{\textbf{Philipp Wintersberger}} is a researcher at TU Wien (Vienna University of Technology). He obtained his doctorate in Engineering Science from Johannes Kepler University Linz specializing on Human-Machine Cooperation. His publications focus on trust in automation, attentive user interfaces, transparency of driving algorithms, as well as UX/acceptance of automated vehicles and have received several awards in the past years. He has co-organized several workshops at CHI and AutomotiveUI, and currently serves as Technical Program Chair for AutomotiveUI’21.

{\textbf{Leigh Clark}} is a Lecturer in Human-Computer Interaction at Swansea University. His research examines the effects of voice and language design on speech interface interactions and how linguistic theories can be implemented and redefined in speech-based HCI. He is co-founder of the CUI conference series and has conducted high impact work in the automotive domain, related to how speech systems can be used in these contexts. 

{\textbf{Daniel Rough}} is a Lecturer in Computing at the University of Dundee. His research focuses on the engagement of users in tailoring software to suit their bespoke requirements, with emphasis on intelligent personal assistants. He was co-organiser of the CUI@IUI workshop.

{\textbf{Philip Doyle}} is a PhD candidate at University College Dublin. His PhD project looks to model people's perceptions of speech interfaces as dialogue partners. Adopting theory and methodological practices from psycholinguistics and cognitive psychology, Philip has published research at CHI, CUI, MobileHCI and IJHCI. He is also a member of the CUI conference steering committee (2017-Present).

{\textbf{Christian Janssen}} is an Assistant Professor in Utrecht University's Department of Experimental Psychology. His research interests are in understanding adaptive human behavior and human-automation interaction through a combination of empirical studies and computational modeling and computer simulations. He has served as General Chair of Auto-UI (2019) and currently serves as Technical Program co-Chair for Auto-UI (2021).

{\textbf{Victoria Banks}} is a Research Fellow in Human Factors Engineering within Engineering and Physical Sciences at the University of Southampton. Her primary research interest centres on agent-based modelling of complex systems operating at varying levels of autonomy and the subsequent validation of these models through simulation and user testing. Victoria has worked closely with Jaguar Land Rover in the design and development of automated driving features using a mix of quantitative and qualitative research methodologies. 

{\textbf{Adam Wyner}} is an Associate Professor in Law and Computer Science at Swansea University, director of the Centre for Innovation and Entrepreneurship in Law an and researches Artificial Intelligence and Law. 

{\textbf{Benjamin Cowan}} is an Associate Professor at University College Dublin’s School of Information \& Communication Studies. His research lies at the juncture between psychology, HCI and computer science in investigating how theoretical perspectives and methods in psychology can used to understand experience and user behaviour in interaction. He is the co-founder of the International Conference on Conversation User Interfaces (CUI) conference series and has run a number of workshops at major HCI conferences (e.g.CHI and Mobile HCI).  
\begin{acks}
This research was conducted with the financial support of the ADAPT SFI Research Centre at University College Dublin. The ADAPT SFI Centre for Digital Content Technology is funded by Science Foundation Ireland through the SFI Research Centres Programme and is co-funded under the European Regional Development Fund (ERDF) through Grant \# 13/RC/2106\textunderscore{}P2.
\end{acks}

\bibliographystyle{ACM-Reference-Format}
\bibliography{AutoUIWorkshop.bib}


\end{document}